\documentclass[aps,nofootinbib,showpacs,showkeys,preprint
tightenlines,twocolumn
] {revtex4}

\usepackage{epsf,epsfig,subfigure,axodraw,graphicx,amsmath,amssymb}
\usepackage{color}

\newcommand{\hfive}{${\bf 5}'$}
\newcommand{\hfiveb}{${\bf\overline 5}'$}
\newcommand{\hten}{${\bf 10}'$}

\newcommand{\five}{${\bf 5}$}
\newcommand{\fiveb}{${\bf\overline 5}$}
\newcommand{\ten}{${\bf 10}$}

 \def\Z{{\bf Z}}

 \def\Z{{\bf Z}}

\def\fivebt{\overline{\bf 5}}
\def\fivet{{\bf 5}}

\def\tent{{\bf 10}}

\begin{document}


\title{\Large\bf Gaugino condensation scale of one family hidden SU(5)$'$, dilaton stabilization and gravitino mass}

\author{Jihn E. Kim\email{jekim@ctp.snu.ac.kr}}
\affiliation{ Department of Physics and Astronomy and Center for Theoretical Physics, Seoul National University, Seoul 151-747, Korea
 }

\begin{abstract}
The hidden SU(5)$'$  with one family, \ten$'$ and \fiveb$'$, breaks supersymmetry dynamically. From the effective Lagrangian approach, we estimate the hidden sector gaugino candensation scale, the dilaton stabilization and the resulting gravitino mass. In some models, this gravitino mass can be smaller than the previous naive estimate. Then, it is possible to raise the SU(5)$'$ confining scale above $10^{13}$ GeV.
\end{abstract}


\maketitle

\section{Introduction}\label{sec:Introduction}

The hidden sector gaugino condensation has long been suggested toward the minimal supersymmetric standard model (MSSM) via the gravity mediation of supersymmetry (SUSY) breaking \cite{Nilles:1982ik}. The gravity mediation of the gaugino condensation introduces the squark mass splitting at the gravitino mass scale which has been naively estimated to be of order $\Lambda^3/M_P^2$, leading to the hidden sector scale $\Lambda$ at $\sim 10^{13}$ GeV.
Including extra dimensions, the Kaluza-Klein modes work $\lq\lq$pro and con" toward a gauge coupling unification below the open-up scale  of the extra dimensions \cite{KimKyaeKK08}. In this regard, it is an interesting attempt to try to obtain a raised hidden sector scale $\Lambda$ together with the gravitino mass more suppressed by $M_P$ \cite{NillesLowen08}. In Ref. \cite{NillesLowen08} the gravitino mass is shown to be highly suppressed, $m_{3/2}\simeq \Lambda^6/M_P^5$, but there a specific form for the superpotential has been assumed.

Thus, in the presence of the hidden sector gaugino condensation, it is an important question to ask, $\lq\lq$What is really the gravitino mass?", and the answer can be applied to the gravity mediation and also to the gauge mediation of SUSY breaking (GMSB) \cite{Kim06GMSUF}. In this paper, we attempt to analyze a proper global SUSY breaking, which is then applied to obtain a gravitino mass in terms of the hidden sector scale. We will obtain a gravitino mass suppressed by $M_P^2$, but a smaller gravitino mass results from the strong dynamics of the hidden sector. The hidden sector dynamics necessarily needs an information how the dilaton $S$ is stabilized. For the dilaton stabilization, we can assume a `race-track' model \cite{racetrack}, but here we will attempt to obtain the dilaton stabilization in the effective Lagrangian approach \cite{VenYank} in the presence of a dynamically generated SUSY breaking source.

The flipped-SU(5) GUT was introduced as another path of SO(10) branching \cite{FlippedSU5}, and recently it has been applied to two dark matter components \cite{HuhKK08}. In  some $\Z_{12-I}$ compactifications of the heterotic string, it is possible to obtain three families in the flipped-SU(5) gauge group in $\Z_{12-I}$ compactifications  \cite{ChoiKimbk,FlipKimKyae,HuhKimKyae09}.

It is known that dynamical breaking of SUSY is possible in some chiral gauge models \cite{DSBSU5,DSBSO10}:
\begin{align}
&\begin{array}{c}
{\rm SU(5)}'\ {\rm with~} \tent'~ {\rm and~} \fivebt'~
\end{array}\\
&\begin{array}{c}
{\rm SO(10)}'\ {\rm with~} {\bf 16}'
\end{array}.
\end{align}
Motivated by this observation, we study one hidden SU(5)$'$ family, \ten$'$ and \fiveb$'$ plus possible $N_f$ numbers of \five$'$ and \fiveb$'$. With one \ten$'$ and one \fiveb$'$, we cannot construct a composite superfield of the form, $\tent\cdot \tent\cdot\tent\cdot\fivebt$. However, there exist two composite superfields constructed with the gluino and matter fields  \cite{Meurice84,Kim06GMSUF},
\begin{eqnarray}
&& Z\sim {\cal W}^\alpha_\beta {\cal W}^\beta_\alpha ,
\label{Zdef}\\
&& Z'\sim\epsilon_{\alpha\gamma\eta\chi\xi} {\cal W}^\alpha_\beta {\cal W}^\gamma_\delta \tent'^{\nu\beta}\fivebt'_{\nu } \tent'^{\eta\delta}\tent'^{\chi\xi},\label{Zpdef}
\end{eqnarray}
where ${\cal W}^\alpha_\beta$ is the hidden sector gluino superfield, satisfying  ${\cal W}^\alpha_\alpha=0, (\alpha=1,2,\cdots,5)$. There is no more SU(5)$'$ invariant independent chiral combination.
The addition of vector-like representations does not change the fate of SUSY breaking of the single family case since the Witten index is not changed \cite{Witten82Index}. The pure SUSY gauge models lead to the effective superpotential  $W\sim Z(\log Z-\rm constant)$ below the confinement scale from which one can always obtain a SUSY condition \cite{VenYank}. On the other hand, one family SU(5)$'$ is classified as an $\lq$un-calculable' model \cite{ShirmanRMP}, but it has been argued that it would break SUSY \cite{DSBSU5,Meurice84}.
\begin{figure}[!]
\vskip 0.5cm
\resizebox{0.45\columnwidth}{!}
{\includegraphics{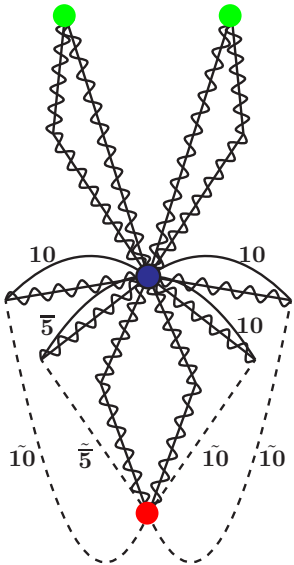}}
\caption{The spider diagram Fig. 5 of Ref. \cite{KimNilles09} shown for the case of $N_f=0$. Two green bullets and a red bullet determine the scale of the instanton effect. }\label{fig:hsu5inst}
\end{figure}
In terms of the composite chiral fields $Z$ and $Z'$, two SUSY conditions cannot be satisfied simultaneously and hence SUSY is broken \cite{KimNilles09}. The key instanton diagram which dictates how one loop effect can be written is shown as a spider diagram in Fig. \ref{fig:hsu5inst}.

\begin{figure}[!]
\vskip 0.5cm
\resizebox{0.7\columnwidth}{!}
{\includegraphics{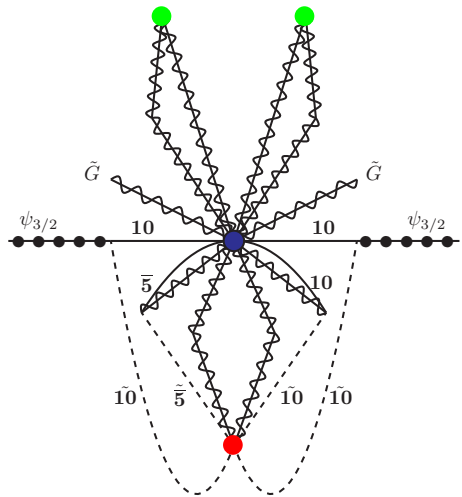}}
\caption{The opened spider diagram of Fig. \ref{fig:hsu5inst}. The gravitino line is the ones with bullets. Other diagrams are also possible by opening other fermion lines by gravitino. Two green bullets and a red bullet determine the scale of the instanton effect, and hence this diagram is proportional to the one given in Fig. \ref{fig:hsu5inst} times $1/M_P^2$.}\label{fig:gravitinomass}
\end{figure}

Assuming that we know the scale of the spider diagram, we can open up some lines of Fig. \ref{fig:hsu5inst} with two gravitinos as shown in Fig. \ref{fig:gravitinomass}. Closing the gluino lines of  Fig. \ref{fig:gravitinomass}, we can estimate the contribution to the gravitino mass by Fig. \ref{fig:gravitinomass} of order
\begin{equation}
m_{3/2}\simeq ({\rm a~ factor})\cdot\frac{f^{' 12}\Phi' \langle\tilde G\tilde G\rangle}{M_P^2\Lambda^{13}}\label{eq:gravmassInt}
\end{equation}
where $\Phi'$ is a dimension 1 chiral field and $f'$ is its decay constant. The expression (\ref{eq:gravmassInt}) has the $1/M_P^2$ and the other factors are calculable in principle if the hidden sector dynamics is known.

In Sec. \ref{sec:Global}, we show how the hidden sector scale $\Lambda$ appears in determining the SUSY breaking minimum. In Sec. \ref{sec:Supergravity}, supergravity effects are included and the gravitino mass is estimated. In the final section, Sec. \ref{sec:Comments}, we comment how our observation can be used to obtain reasonable MSSM models from an ultra-violet completed theory.

\section{Global SUSY breaking in Chiral SU(5)$'$ gauge theory} \label{sec:Global}

Now, let us proceed to consider the one family SU(5)$'$ model, with \ten$'$ and \fiveb$'$. Then, the following SU(5)$'$ singlet vector fields can be considered,
\begin{eqnarray}
V_{10}\sim \tent'\cdot \tent'^* ,
\quad V_{\bar 5}\sim \fivebt'\cdot \fivebt'^*,
\quad V_5\sim \fivet'\cdot \fivet'^*.\nonumber\label{Vsdef}
\end{eqnarray}
These vector fields do not contribute to an effective potential but can give rise to couplings of the form
\begin{equation}
\int d^4\theta [V_{10} g_{10}+ V_{\bar 5} g_{\bar 5}+ V_{5} g_{5}+\cdots]\nonumber
\end{equation}
where $g_{10},g_{\bar 5}$ and $g_5$ are real functions of the SU(5)$'$ singlet composite fields. Since we lack a method to treat this general form, we restrict to the known nonperturbative effects guided by the instanton interaction. Then, we can consider two SU(5)$'$ singlet composite chiral fields considered in Eqs. (\ref{Zdef}) and (\ref{Zpdef}).

Due to the index theorem, the model with a \hten, a \hfiveb, plus $N_f$ copies of \hfive\ and \hfiveb\ breaks SUSY. The quantum numbers of the global symmetries are as shown in Table \ref{tab:SUfive} \cite{KimNilles09}. We defined the charges such that the U(1)$_R$ is anomaly free, and there is only one anomalous U(1): the U(1)$_A$. The U(1)$_A$ is broken by the SU(5)$'$ instantons. Below the SU(5)$'$ confinement scale, we can consider only two effective chiral fields as suggested in \cite{Meurice84}. In addition, we must consider the global symmetries of the flavors \hfive\ and \hfiveb: ${\rm SU}(N_f)\times {\rm SU}(N_f+1)$. This flavor symmetry $SU(N_f)\times SU(N_f+1)$ must be realized below the confinement scale $\Lambda$ as \cite{tHooftnat}:

\begin{table}
\begin{center}
\begin{tabular}{c|cccc}
&~ $U(1)_A$& $U(1)_B$& $U(1)_R$~ & $U(1)_q$ \\
\hline\\[-0.8em]  \hten~ & $\frac32$&$1$&~ $\frac29(N_f-10)$ &  $-\frac13(2N_f+1)$\\[0.3em]
 \hfiveb~ & $1$&$-3$~& $\frac23$ &  $1$\\[0.3em]
 \hfive$^i$~ & $1$&$3$& $\frac23$ &  $1$\\[0.3em]
 \hfiveb$_i$~ & $1$&$-3$~& $\frac23$ &  $1$\\[0.3em]
 $Z$ &0&$0$&2 &  $0$\\[0.3em]
 $Z'$ &$\frac{11}{2}$&$0$&~$\frac23(N_f-6)$ &  $-2N_f$\\[0.3em]
 ${\cal M}$ &$2N_f$&$0$&$\frac43N_f$ &  $2N_f$\\[0.3em]
 $Z'_{\cal M}$ &$2N_f+\frac{11}{2}$&$0$&$2(N_f-2)$ &  $0$\\[0.3em]
 $\Lambda^{3N_c-2-N_f}$ &~$2N_f+\frac{11}{2}$&$0$&0 &  $0$
\end{tabular}
\end{center}
\caption{Here, $Z=\tilde G\tilde G$, $Z'={\cal WW}\tent'\tent'\tent'\fivebt'$, and ${\cal M}= ${\rm Det.}\hfive$^i$\hfiveb$_j$.} \label{tab:SUfive}
\end{table}

\begin{itemize}
\item[$(a)$] If the flavor symmetry remains unbroken, there appear $N_f$ singlets for the fundamental of $SU(N_f)$ and $N_f+1$ singlets for the fundamental of $SU(N_f+1)$.
\item[$(b)$] If all or a part of the flavor symmetry is broken, there appear the corresponding composite Goldstone boson multiplets.
\end{itemize}
For $(a)$, we cannot find the matching number for the composite singlets. For $(b)$, we can find the matching number for the Goldstone bosons, i.e. in the phases of
\begin{eqnarray}
&&Z'_{(i)}\sim \epsilon_{acfgh}{\tilde G}^{a}_b{\tilde G}^c_{d} \tent^{eb}\fivebt_{e(i)} \tent^{fd}\tent^{gh},\nonumber\\
&&\hskip 1cm \quad i=1,\cdots, N_f+1,
\label{ZpGold}\\
&& {\cal M}\sim \fivet^{(i)}\fivebt_{(j)},\quad {\rm Tr}{\cal M}=0,\nonumber\\
&&\hskip 1cm \quad i=1,\cdots, N_f,\ j=1,\cdots, N_f+1.\label{MGold}
\end{eqnarray}
The number of Goldstone bosons in (\ref{ZpGold}) is $N_f+1$, and the number of Goldstone bosons in (\ref{MGold}) is $N_f(N_f+1)-1$. Thus, the total number of Goldstone bosons for the case of ($b$) is $(N_f^2+2N_f)$. This is the number resulting if $SU(N_f)\times SU(N_f+1)$ is broken down to $SU(N_f)$: $(N_f^2-1)+((N_f+1)^2-1)-[N_f^2-1]\Rightarrow N_f^2+2N_f$.


So we consider the composite fields considered in (\ref{Zdef}) and (\ref{Zpdef}) with $N_f=0$. The instanton interaction is a determinant, i.e. the flavor group singlet, and we consider all these fields appearing in the diagram, like $Z'_{\cal M}$ for $N_f\ne 0$. Then, the effective superpotential is
\begin{equation}
W_{\rm SU(5)}=Z\left[\log\left(\frac{Z^{2-N_f} Z'_{\cal M}}{\Lambda^{3N_c-2-N_f}} \right) -\alpha\right].\label{eq:SUfW}
\end{equation}
Since there appear only one combination in (\ref{eq:SUfW}), we can redefine $Z'_{\cal M}$ for the effective interaction as considered in Table \ref{tab:SUfive}. But certainly there are more fields if $N_f\ne 0$, for which the stabilization of these extra ($N_f^2+2N_f-1$) fields must be taken into account also, which is out of scope of this paper.

Toward the directions $Z$ and $Z'_{\cal M}$, for the dynamically generated effective superpotential respecting the global symmetries, we use Eq. (\ref{eq:SUfW}) \cite{KimNilles09} where $\alpha$ is considered as a coupling. It was shown that for $N_f=3$, the SUSY conditions cannot be satisfied and SUSY is dynamically broken \cite{KimNilles09}.
Let us define the following $\Phi$ and $\Phi'$ such that their engineering dimensions are 1,
\begin{equation}
\Phi=Z/f^2,\quad \Phi'=Z'_{\cal M}/f^{'6+2N_f}
\end{equation}
where $f$ and $f'$ are scale parameters. In principle, the decay constants $f$ and $f'$ are determined by the hidden sector dynamics. These are expected to be near the scale $\Lambda$.
The effective superpotential is
\begin{align}
W_{\rm SU(5)}&=f^2\Phi\left[\log\left(\frac{f^{4-2N_f} f^{' 6+2N_f}
\Phi^{2-N_f} \Phi'}{\Lambda^{13-N_f}} \right) -\alpha\right]\nonumber\\
&=f^2\Phi\left[\log\left(\frac{f^{4-2N_f} f^{' 6+2N_f}
\Phi^{2-N_f} \Phi'}{\Lambda^{13-N_f}e^{\alpha+1}} \right) +1\right]\nonumber
\end{align}
Let us rescale $\Lambda$ such that $\Lambda^{13-N_f} e^{\alpha+1}\to \Lambda^{13-N_f}$ and hence we can use the following superpotential without loss of generality,
\begin{align}
W_{\rm SU(5)}
&=f^2\Phi\left[\log\left(\frac{f^{4-2N_f} f^{' 6+2N_f}
\Phi^{2-N_f} \Phi'}{\Lambda^{13-N_f}} \right) +1\right].\label{eq:NormNf}
\end{align}
Consider the case of $N_f=3$ which leads to a simple analysis,
\begin{equation}
W_{\rm SU(5)}=f^2\Phi\left[\log\left(\frac{f^{' 12} \Phi'}{f^2\Lambda^{10}\Phi} \right) +1\right]
\label{eq:Norm}
\end{equation}
\begin{equation}
\begin{array}{c}
\frac{\partial W_{\rm SU(5)}}{\partial \Phi} =f^2\left[\log\frac{\Phi'}{\Phi}+\log\left(\frac{f^{' 12} }{f^2\Lambda^{10}} \right) \right]\\
\frac{\partial W_{\rm SU(5)}}{\partial \Phi'} =f^2\frac{\Phi}{\Phi'}\label{eq:Fterms}
\end{array}
\end{equation}
The SUSY conditions from the upper and lower equations of (\ref{eq:Fterms}) are $\Phi'/\Phi=   f^2\Lambda^{10}/f^{' 12}$ and  $\Phi/\Phi'=  0$, respectively, which are mutually inconsistent.

Note that $\Phi$ and $\Phi'$ possess 4 real fields which are denoted as $\rho, \rho', \delta,$
and $\theta$,
\begin{align}
\Phi=\rho e^{i\delta},\quad \Phi'=\rho' e^{i(\delta+\theta)}.
\end{align}
We can write $V_0$ as
\begin{align}
V_0/f^4&=\left|\log\frac{\Phi'}{\Phi}+\log\left(\frac{f'^{12} }{f^2\Lambda^{10}} \right) \right|^2+\left|\frac{\Phi}{\Phi'}\right|^2\nonumber\\
 &=\left(-\log\xi+\log\epsilon \right)^2+\xi^2+\theta^2\label{eq:V}
\end{align}
where
\begin{equation}
\xi=\frac{\rho}{\rho'},\quad\epsilon=\frac{f'^{12} }{f^2\Lambda^{10}} .\label{eq:rho}
\end{equation}

The minimum conditions $\frac{\partial{V}_0}{\partial\rho}=0$ and $\frac{\partial{V}_0}{\partial\rho'}=0$ give the same condition
\begin{align}
\log\xi =\log\frac{f'^{12} }{f^2\Lambda^{10}}-\xi^2.  \label{eq:rhopxicond}
\end{align}

\begin{figure}[!]
\vskip 0.5cm
\resizebox{0.8\columnwidth}{!}
{\includegraphics{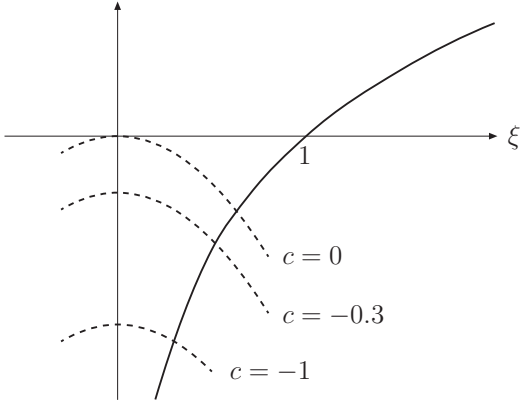}}
\caption{A schematic view of the solution of Eq. (\ref{eq:rhopxicond}). The vertical axis is the LHS or the RHS of Eq. (\ref{eq:rhopxicond}). The solid curve is the LHS and dashed curves are the RHS. Here, $c$ represents $\log\frac{f'^{12} }{f^2\Lambda^{10}e^{(\alpha+1)}}$. For a large negative $c$, the solution for $\xi$ is exponentially small.}\label{fig:Solution}
\end{figure}

For given $f$ and $f'$, one can find a solution as depicted in Fig. \ref{fig:Solution}. Without SUSY, we expect that the condensation of two gluinos is much more stronger than that of hidden sector quark and anti-quark pairs. For the chiral representations, we have less information on the condensations. Furthermore, SUSY can change this view of condensations of fermions, and therefore we treat $f$ and $f'$ as unknown free parameters near but somewhat below the scale $\Lambda$. Then, the RHS of Eq. (\ref{eq:rhopxicond}) is expected to be negative, and a solution for $\xi<1$ is a possibility. Let that number be for example O(--10). Then, $\xi\sim e^{-10}\simeq 4.5\times 10^{-5}$. The h-gluino condensation scale is $f^2\langle\rho \rangle$ which in this case is of order $(3.6\times 10^{-2}\Lambda)^3$ which gives about two orders smaller gaugino condensation scale compared to $\Lambda$. For example, for $\alpha=-1$ and $f=\Lambda$ we have
$$
\log\xi =-\xi^2+12\log\frac{f'}{\Lambda}
$$
and find an approximate solution $\xi\simeq (f'/\Lambda)^{12}$ or $\rho\simeq \rho'(f'/\Lambda)^{12}$. So, the gaugino condensation scale is of order
\begin{equation}
\langle \tilde G\tilde G\rangle^{1/3}\simeq f^{2/3}\langle\rho'\rangle^{1/3}\left(\frac{f'}{\Lambda}\right)^{4}\sim \left(\frac{\langle\rho'\rangle}{\Lambda}\right)^{1/3}
\left(\frac{f'}{\Lambda}\right)^{4}\Lambda
\end{equation}
The power $\frac13$ of $\langle\rho'\rangle$ is small, and hence its effect is minor compared to the effect of the power of $f'$. So, the gaugino condensation scale can be exponentially smaller than the hidden sector scale $\Lambda$ for $f'/\Lambda\ll 1$, which is traced to the property of the high engineering dimension of the operator $Z'$.

\section{Supergravity and gravitino mass}\label{sec:Supergravity}

If supergravity is considered, the above DSB leads to a nonvanishing gravitino mass. The gravitino mass is given in the supergravity Lagrangian as \cite{DZ77, Cremmer83,Nilles84},
\begin{eqnarray}
m_{3/2}=\frac{M_S^2}{\sqrt3 M_P} .\label{eq:gravmass}
\end{eqnarray}
In our case, the SUSY breaking parameter $M_S^2$ is the F-terms,
\begin{eqnarray}
M_S^2=\sum_i F_i=f^{\alpha\beta}_\kappa (G^{-1})^\kappa_i\lambda^\alpha \lambda^\beta +e^{G/2} (G^{-1})^j_i G_j\label{eq:gravSUSYbr}
\end{eqnarray}
where $G$ is the superpotential modified K\"ahler function, $G=K +\log|W|^2$,  $f^{\alpha\beta}$ is the gauge kinetic function, and $f^{\alpha\beta}_\kappa $ is the derivative of $f^{\alpha\beta}$ with respect to the chiral field $\phi^\kappa$. If the gauge kinetic function is a constant, the F-terms of $Z$ and $Z'$ give the gravitino mass. Since the effective Lagrangian is arising from Fig. \ref{fig:hsu5inst}, the gravitino mass can be shown as Fig. \ref{fig:gravitinomass}.

Let us now include the dilaton superfield $S$ in the gauge kinetic function.  Motivated by the heterotic string, let us consider the following K\"ahler potential,
\begin{equation}
K(S,S^*;\Phi,\Phi^*,\Phi',\Phi'^*)= -M_P^2\log(S+S^*)+ \Phi\Phi^*+ \Phi'\Phi'^*,
\end{equation}
from which the potential is calculated as
\begin{equation}
V=e^{K/M_P^2}\left[ (K^{-1})^{ij} D_i W\overline{D_j} \overline{W}-\frac{3}{M_P^2}| W|^2\right]
\end{equation}
where $D_i =(\partial/\partial \phi^i) +(\partial K/\partial \phi^i) $ and $\phi_i=\{S,\Phi,\Phi'\}$. Considering the result of Sec. \ref{sec:Global}, let us take
\begin{eqnarray}
&W\equiv Z\left[\log\left(\frac{Z^{2-N_f}  Z'_\Phi}{\Lambda^{3N_c-2-N_f}} \right) -\alpha-\frac{S}{4M_P}\right]+C\nonumber\\
&=W_0-\frac{S}{4M_P}Z +C
\label{eq:Wsugra}
\end{eqnarray}
where  $W_0$ is $W_{\rm SU(5)}$ of Eq. (\ref{eq:SUfW}) with $\alpha$ modified by the dilaton coupling, and a constant $C$ is added as a free parameter resulting from the U(1)$_R$ breaking gravitational interaction. The term $-(1/4M_P) SZ$ in $W$ has been considered previously \cite{Bailin99}, but its consequence on the gravitino mass from one family SU(5)$'$ model has never been presented before. Since we are looking for a solution at a large value of $S$, i.e. in the perturbative region, we can just read off the original gaugino coupling below the confinement scale as $-\frac{1}{4M_P} SZ$. Then, we must consider the following K\"ahler potential dependence,
\begin{align}
& e^{K}=\frac{1}{S+S^*}e^{(|\Phi|^2+|\Phi'|^2)} \nonumber\\
&K_{SS^*}=\frac{1}{(S+S^*)^2},\ K_{\Phi\Phi^*}=1,\ K_{\Phi'\Phi'^*}=1 \nonumber\\
&D_SW=\frac{-1}{S+S^*}\left[W_0+C -\frac{f^2 S\Phi}{4} +\frac{f^2 (S+S^*)\Phi}{4}\right]
\nonumber\\
&D_\Phi W 
=\frac{\partial W_0}{\partial\Phi} -\frac{f^2 S}{4}+(W_0+C-\frac{f^2 S\Phi}{4})\Phi^*
\\
&D_{\Phi'} W=\frac{\partial W}{\partial\Phi'}+ \frac{\partial K}{\partial\Phi'}W= \frac{\partial W_0}{\partial\Phi'}+(W_0+C-\frac{f^2 S\Phi}{4})\Phi'^*
\nonumber
\end{align}
so that we obtain
\begin{widetext}
\begin{align}
& K^{-1~SS^*}D_SW(D_SW)^*=\left|W_0+C-\frac{f^2 S\Phi}{4} +\frac{f^2 (S+S^*)\Phi}{4}\right|^2
\nonumber\\
&|D_\Phi W|^2=\left|\frac{\partial W_0}{\partial\Phi} -\frac{f^2 S}{4}+(W_0+C-\frac{f^2 S\Phi}{4})\Phi^*\right|^2
\nonumber\\
&|D_{\Phi'} W|^2=\left|\frac{\partial W_0}{\partial\Phi'} +(W_0+C-\frac{f^2 S\Phi}{4})\Phi'^*\right|^2 \label{eq:WwithC}
\end{align}
where we set $M_P= 1$ for a moment. Thus, the potential is expressed as
\begin{align}
V&=\frac{e^{(|\Phi|^2+|\Phi'|^2)/M_P^2}}{(S+S^*)/M_P}\Big\{V_0+
\frac{|W_0-\frac{1}{4M_P}f^2S\Phi|^2}{M_P^2}\left(\frac{|\Phi|^2}{M_P^2}+
\frac{|\Phi'|^2}{M_P^2}-2\right) \nonumber\\ &+\frac{f^4}{16}(\frac{|S|^2}{M_P^2}+\frac{(S+S^*)^2|\Phi|^2}{M_P^4})
+\left[-\frac{f^2S^*}{4M_P}  \left(\frac{\partial W_0}{\partial \Phi} +(W_0-\frac{f^2\Phi}{4M_P}S)\frac{\Phi^*}{M_P^2}\right)+{\rm h.c.}\right]
\nonumber\\
&+\left[(\frac{W_0^*}{M_P^2}-\frac{f^2S^*\Phi^*}{4M_P^3}) \left(\frac{\partial W_0}{\partial \Phi}\Phi+\frac{\partial W_0}{\partial \Phi'}\Phi'+\frac{1}{4M_P} f^2(S+S^*)\Phi\right)+{\rm h.c.}\right]\Big\}
\label{eq:VnoC}
\end{align}
\end{widetext}
where we omitted the terms depending on $C$ of Eq. (\ref{eq:WwithC}). This $C$ will be used only for adjusting the cosmological constant to zero, and hence for simplicity we will neglect its dependence of the dilaton stabilization.

Let us choose the real fields as
\begin{equation}
\rho, \rho', \delta, \theta, \ {\rm and~} S=\sigma e^{i\theta_{MI}},
\end{equation}
where the following $2\pi$ ranges of the angles are assigned,
\begin{equation}
\theta=(-2\pi,0],\quad\delta=(-\pi,\pi],\quad \theta_{MI}=(-\pi,\pi].
\end{equation}
Since the principal value of $\theta$ appears as polynomials in our expression, we choose our convenient range of $\theta$ ending at 0.
Then, we obtain the following potential
\begin{widetext}
\begin{align}
\frac{2\sigma_{cr}/M_P}{e^{(\rho^2+\rho^{'2})/M_P^2}} &V=V_0
+\frac{f^4\rho^2}{M_P^2}\left(\frac{\rho^2}{M_P^2}+ \frac{\rho^{'2}}{M_P^2}-2\right)\cdot
\Big(|-\log\xi+\log\epsilon+1|^2 + \theta^2
+\frac{\sigma^2}{16 M_P^2}\nonumber\\
&-\frac{\sigma_{cr}}{2M_P}(-\log\xi +\log\epsilon+1) +\frac{\sigma_{cr}}{2M_P}\theta\tan\theta_{MI} \Big)
+\frac{f^4\sigma^2}{16M_P^2}\Big(1+ \frac{4\rho^2\cos^2\theta_{MI}}{M_P^2} \Big)\nonumber\\
&-\frac{f^4\sigma_{cr}}{2M_P}(1+\frac{\rho^2}{M_P^2}) \left[(-\log\xi+\log\epsilon)+\theta\tan\theta_{MI} \right]
-\frac{f^4\rho^2\sigma_{cr}}{2M_P^3} +\frac{f^4\rho^2\sigma^2}{8M_P^4}\nonumber\\
&+\frac{2f^4\rho^2}{M_P^2}\Big\{( -\log\xi+\log\epsilon+1-\frac{\sigma_{cr}}{4M_P} )( -\log\xi+\log\epsilon+1+\frac{\sigma_{cr}}{2M_P} ) \nonumber\\
&+\theta(\theta-\frac{\sigma\sin\theta_{MI}}{4M_P}) \Big\}
\end{align}
where $V_0$ is given in Eq. (\ref{eq:V}). In applying the above equation, we take the positive parameters except the angles. Representing $V$ in terms of dimensionless fields, $\tilde\sigma=\sigma/M_P, h=\rho/M_P$ and $h'=\rho'/M_P$, we obtain
\begin{align}
 \frac{V}{f^4}=&\frac{e^{(h^2+h^{'2})}}{2\tilde\sigma\cos\theta_{MI}} \Big\{\frac{V_0}{f^4}
-\frac{\tilde\sigma}{2}\Big[(-\log\xi +\log\epsilon)\cos\theta_{MI}+\theta\sin\theta_{MI}\Big]
+\frac{\tilde\sigma^2}{16}\nonumber\\
&\quad\quad\quad\quad\quad+h^2 \Big[\tilde\sigma\cos\theta_{MI}( -\log\xi+\log\epsilon+1)+ \frac{\tilde\sigma^2}{4} \cos^2\theta_{MI}\Big]\label{eq:finV}\\
&+{h^2}\left({h^2}+ h^{'2}\right)\cdot
\Big[(-\log\xi+\log\epsilon+1)^2
-\frac{\tilde\sigma}{2}\cos\theta_{MI}(-\log\xi+\log\epsilon+1)\nonumber\\
&\quad\quad\quad\quad\quad +
\frac{\tilde\sigma^2}{16 }-\frac{\tilde\sigma}{2} \theta \sin\theta_{MI}+ \theta^2\Big]\Big\}.\nonumber
\end{align}
Note that in the limit of $M_P\to\infty$, i.e. $h,h'\to 0$, the dilaton is stabilized at
\begin{equation}
\tilde\sigma\simeq 4\sqrt{V_0/f^4} = 4\sqrt{(-\log\xi+\log\epsilon)^2+\xi^2}.
\end{equation}
\end{widetext}

\begin{figure}[!]
\vskip 0.5cm
\resizebox{0.7\columnwidth}{!}
{\includegraphics{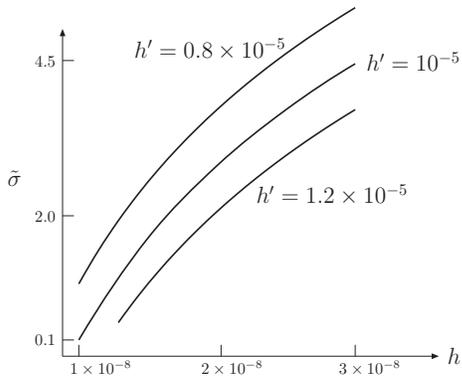}}
\caption{The dilaton stabilization point $\tilde\sigma=
|S|/M_P$ as a function of $h$. We take $f=\Lambda/2$ and $f'=\Lambda/2$.}\label{fig:Dilastab}
\end{figure}

In Fig. \ref{fig:Dilastab} we show the dilaton stabilization for $f=f'=\Lambda/2$. At this dilaton stabilized point, the potential is $V\simeq (1/\tilde\sigma)\sqrt{V_0}[\sqrt{V_0}-f^2\log(\epsilon/\xi)]$. So far we neglected $C$ of Eq. (\ref{eq:Wsugra}), and $V$ is tuned to 0 by an appropriate choice of $C$.

In the limit $M_P\to\infty$, we obtain $\xi\to \epsilon$ for a small $\epsilon$. Then, the F-terms of $S,\Phi$ and $\Phi'$ fields at the minimum of the supergravity potential are, viz. Eq. (\ref{eq:Fterms}),
\begin{align}
&F_S\simeq -\frac{f^2 he^{i\delta}}{4}, F_\Phi'\simeq f^2\xi e^{-i\theta},\nonumber\\
& F_\Phi\simeq f^2(-\log\xi+\log\epsilon+i\theta),
\end{align}
so that the gravitino mass from Eq. (\ref{eq:gravmass}) becomes
\begin{align}
m_{3/2} &\simeq \frac{f^2}{\sqrt3 M_P}\Big|-\frac{he^{i\delta}}{4}-\log\xi+\log\epsilon\nonumber\\
&\quad\quad\quad\quad\quad+\xi e^{-i\theta} \quad+i\theta\Big|\nonumber\\
&\simeq\frac{f^2}{\sqrt3 M_P}|(-\log\xi+\log\epsilon)^2+\theta^2|^{1/2}\label{eq:tunedgrm}
\end{align}
where we used the limit $\xi\to 0$ and $h\to 0$ in the second line. Toward a suppressed gravitino mass, we need both the ratio $\epsilon/\xi$ being sufficiently close to 1 and $\theta\simeq 0$. Then, the gravitino mass is accordingly suppressed. From (\ref{eq:finV}), in this limit the phase stabilization is determined at $\theta=0$ and $\cos\theta_{MI}=1$. At this vacuum, Eq. (\ref{eq:tunedgrm}) can give a reduced gravitino mass for $\xi/\epsilon\simeq 1$.

Thus, in principle, the gravitino mass is determined by the effective Lagrangian approach, with which the dilaton is stabilized. But due to our ignorance on the parameters in the confining sector, $f,f'$, and $K$, we cannot determine the gravitino mass exactly in the effective Lagrangian approach.

Because of the difficulty in estimating the order of the absolute magnitude, one may resort to the original dynamical source due to the instanton diagram. Fig. \ref{fig:hsu5inst} has two more loops (the hidden sector gluino--quark loops) compared to Fig. \ref{fig:gravitinomass} (the $N_f=0$ case). A naive estimation of these two loops would be $(1/8\pi^2)^2$ times the hidden sector coupling  times the relevant mass scale of the hidden sector. Since Fig. \ref{fig:hsu5inst} has 11 loops, the momenta going around one loop is averaged to $\Lambda/11$. So, we roughly estimate the missing two loops contribute $\sim (1/64\pi^4)(\Lambda/10)^\nu$ where $\nu$ can be taken as 4, the hidden sector coupling of O(1) and we use 10 instead of 11 (considering Fig. \ref{fig:hsu5inst}) or 9  (considering Fig. \ref{fig:gravitinomass}). If there are $N_f$ pairs of $\fivet'$ and $\fivebt'$, we may divide $\Lambda$ by $10+N_f$. Therefore, we estimate the diagram shown in  Fig. \ref{fig:gravitinomass} as
\begin{equation}
 \left(\frac{64\pi^4(10+N_f)^4}{\Lambda^4}\right)\frac{f^{' 12+2N_f}\Phi'}{M_P^2K^{9+2N_f}}\bar\psi_{3/2} \psi_{3/2}\tilde G\tilde G \end{equation}
from which we estimate for $K\simeq\Lambda$
\begin{equation}
m_{3/2}\simeq 64\pi^4(10+N_f)^4~\frac{f^{' 12+2N_f}\langle\Phi'\rangle \langle\tilde G\tilde G\rangle}{M_P^2 \Lambda^{13+2N_f}}.\label{eq:massgravti}
\end{equation}
For example, taking $\langle\tilde G\tilde G\rangle/K^3\sim \langle\tilde G\tilde G\rangle/\Lambda^3\sim 2.5\times 10^{-7}$ due to small $\langle\Phi\rangle$, $\Phi'=\Lambda, f\sim 1/2,  f'\sim 1/10$, and $\Lambda=10^{16}$ GeV, Eq. (\ref{eq:massgravti}) gives $3.4\times 10^{-2}\Lambda^3/M_P^2\sim 7.5\times 10^{-6}$ GeV for $N_f=3$. However, we have observed that the gravitino mass has one inverse power of $M_P$ and hence, correcting the above number by multiplying $M_P/F^{1/2}\sim 10^5$, we obtain $m_{3/2}\sim 1$ GeV. Therefore, hidden sector dynamics may lead to a very close $\xi$ and $\epsilon$ in Eq. (\ref{eq:tunedgrm}).

\section{Conclusion and Comments}
\label{sec:Comments}

In this paper, we estimated the gravitino mass in one hidden family SU(5)$'$ models in terms of the hidden sector scale $\Lambda$ in the vacuum where the dilaton is stabilized and showed that it is possible to reduce the gravitino mass than the previous naive estimate. So, starting with the hidden sector coupling much above $\Lambda\gg 10^{13}$ GeV, one can obtain a sufficiently small gravitino mass or the sub-TeV mass splittings in the visible sector superfields by gravity mediation. Then, a TeV order SUSY scale can be in principle calculated in the GMSB or/and anomaly mediation. The reduced gravity mediation effect can be included if it is non-negligible.

For an ultra-violet completion of this kind of DSB with one hidden family SU(5)$'$ group, compactifications in  $\Z_{12-I}$ orbifolds in the heterotic string and F-theory are suitable. For example, we find
in $\Z_{12-I}$ orbifold compactifications that there frequently  appear hidden sector one family SU(5)$'$ groups. Among these, choosing three visible sector families, the number of such models is drastically reduced. Requiring other phenomenological constraints, this number is further reduced. We presented two such models before \cite{Kim06GMSUF,HuhKimKyae09}. If one starts from a universal gauge coupling at the GUT unification point ($\gtrsim 10^{16}$ GeV) of the visible sector couplings, both of these models predict the hidden sector scale above $10^{13}$ GeV, naively predicting the gravitino mass much above the TeV scale. Of course, contributions from the Kaluza-Klein modes between the string scale and the GUT scale may change \cite{KimKyaeKK08} this undesirable feature, but the models presented in \cite{Kim06GMSUF,HuhKimKyae09} and probably most models listed in \cite{KimKyaeKK08}  work in the opposite direction because the $\beta$ function of the hidden sector gauge group is smaller than that of the visible sector gauge group. The mechanism we discussed in this paper can remedy this dilemma. Namely, starting with a universal gauge coupling at the GUT unification point even with a large hidden sector scale, presumably near $\Lambda \sim 10^{16}$ GeV, one can achieve a sufficiently small gravitino mass so that a TeV order visible sector SUSY scale may result from the other sources such as from the GMSB \cite{GMSB99}.

\acknowledgments{ I thank J.-H. Huh and B. Kyae for valuable comments, and especially H. P. Nilles for discussions in the initial stage of this work.
 This work is supported in part by the Korea Research Foundation, Grant No. KRF-2005-084-C00001.
}
\vskip 0.5cm

\appendix\centerline{\bf Appendix}\vskip 0.2cm

In this Appendix, we present the mixing between the spin $\frac32$ component and the spin $\frac12$ component by the DSB. We note that Fig. \ref{fig:gravitinomass} is the essential one leading to the gravitino mass, including both the source of DSB of Fig. \ref{fig:hsu5inst}  and the U(1)$_R$ invariance: The DSB source, the green and red blobs, give the U(1)$_R$ invariance and the U(1)$_R$ charges of two gravitino lines, and two gluino lines add up to zero.  So, Fig. \ref{fig:gravitinomass} is the basis for calculating the gravitino mass, by relating the spin-$\frac32$ component to the Goldstino component spin-$\frac12$. It is in parallel to the gauge boson mass by coupling the spin-1 component $A_\mu$ to the Goldstone boson component spin-0 $a$. In the flat space, we note that for a particle with a nonzero helicity, the mass term changes the helicity $h$ because the massive particle cannot move faster than the speed of light and one can Lorentz-boost such that the helicity $h$ looks changed. So, the coupling of two helicity states with $\Delta h=1$ is the mass of the nonzero helicity particle.

For spontaneous breaking of U(1) gauge symmetry, one starts with a coupling $e(\partial_\mu\phi^*+eA^\mu \phi^*) \phi A_\mu$ from $(eA^\mu \phi^*)(e \phi A_\mu)$ where $\phi^*$ part is shown as a gauge invariant form under $A^\mu\to A^\mu -(i/e)\partial^\mu\lambda(x)$ and $\phi^*\to e^{i\lambda(x)}\phi^*$. If $\langle\phi\rangle=0$, the U(1) gauge boson cannot couple to a longitudinal component. However, if the gauge symmetry is broken by a vacuum expectation value of $\phi$, one can take a unitary gauge to represent $\phi$ as a Goldstone boson dependent function $\phi=(v/\sqrt2)e^{ia/f}$. Then, the quantity inside the bracket $(eA^\mu \phi^*)$ can be gauge transformed to $A^\mu\to A^\mu -(i/e)\partial^\mu\lambda(x)$ and $\phi^*\to e^{i\lambda(x)}\phi^*$. The $x$ dependent function $a(x)/f$ is identified as the dimensionless gauge function $\lambda(x)$, and one obtains the $A_\mu$ (transverse component) to $\partial_\mu\lambda$ (longitudinal component = Goldstone boson) coupling $e(ev\partial^\mu \lambda/\sqrt2) (v/\sqrt2)A_\mu$, and hence the gauge boson mass is $e^2v^2$.

We can view Fig. \ref{fig:gravitinomass} as the SUSY invariant coupling respecting the global symmetries, and close the gluino lines to introduce another green blob. Namely, we integrate out strongly interacting fields and consider only light fields $Z$ and $Z'_\Phi$, and hence only the second term of Eq. (\ref{eq:gravSUSYbr}) is considered. Instead of  Fig. \ref{fig:gravitinomass}, along the above paragraph we try to obtain the $\psi_{3/2}$ (transverse component) to $\partial_\mu\lambda_{1/2}$ (Goldstino) coupling to obtain the gravitino mass where $\lambda_{1/2}$ is defined to carry dimension $\frac12$. Namely, $\lambda_{1/2}=(M_P/F)\psi_G$ where $\psi_G$ is the Goldstino field.

In the zero vacuum energy, let us consider the magnetic moment type gravitino coupling to chiral fields to $W$ \cite{Cremmer83,Nilles84} and the Goldstino ($\psi_G$) coupling
\begin{equation}
e^{-1}{\cal L}_F\to \left\{
\begin{array}{l}
M_P e^{G/2}\psi_\mu\sigma^{\mu\nu}\psi_\nu =\frac{W}{M_P^2}\psi_\mu\sigma^{\mu\nu}\psi_\nu\\
 XGG
 \end{array}\right.\label{eq:gravGold}
\end{equation}
where we can use (\ref{eq:NormNf}) for $W$, $G$ is the Goldstino superfield, and $X$ is an auxiliary field splitting scalar partner of $G$ from the massless $\psi_G$. So, we can assign the auxiliary field $X$ as
\begin{equation}
X=F\vartheta^2\label{eq:Fauxill}
\end{equation}
so that the Goldstino $\psi_G$ remains massless in the broken SUSY case. $\psi_G$ has dimension 3/2.
The first term of (\ref{eq:gravGold}) is an interaction term and is not a gravitino mass term yet.  The SUSY transformations of (\ref{eq:gravGold})  relate spin-$\frac32$ components to spin-$\frac12$ components. Firstly, the SUSY transformation of $\psi_\mu$ to $\psi_{1/2} $ generates $(3W/M_P^2)\psi_{1/2} \psi_{1/2} $ because three components in $\psi_\mu$ in the $\gamma^\mu\psi_\mu=0$ gauge goes to one component in $\psi_{1/2}$. Note that $\psi_{1/2}$ has dimension 3/2 and the suffix $\frac12$ denotes a two-component spinor. Therefore, due to the Goldstino definition as the F-term breaking of SUSY as given in Eq. (\ref{eq:Fauxill}), we interpret $\psi_{1/2}=\psi_G/\sqrt3$. Dimension $1/2$ Goldstino field $\lambda_{1/2}$ is the Goldstino direction $\lambda_G=(M_P/F)\psi_G$, taking into account the gravitational charge $1/M_P$ and the SUSY breaking scale $F$. Dimension $-1/2$ Goldstino field $\lambda_{-1/2}$ is required to have the same dimension as the Grassmann variable $\vartheta$ and hence $\lambda_{-1/2}=(M_P/F)\lambda_{1/2}$. The downward (twice) SUSY transformations of $\psi_\mu$ will lead to $\partial_\mu(\lambda_{-1/2}/\sqrt3)$. The upward (twice) SUSY transformations of a chiral field in $W$ will lead to the F-term of that field, i.e. $F_\Phi$ for example. Thus, the upward SUSY transformation of $(W/M_P^2)$ will be
\begin{align}
 \sum_{\phi_\kappa=\Phi,\Phi'} \frac{\partial W}{\partial \phi_\kappa}\frac{F_{\phi_\kappa}}{M_P^2}&= \frac{f^2}{M_P^2}[(-\log\xi+\log\epsilon)F_{\Phi}+\xi F_{\Phi'}]\nonumber\\
=\frac{F^2}{M_P^2}&=\frac{f^4}{M_P^2}[(-\log\xi+\log\epsilon)^2+\xi^2 ]
\end{align}
where we used Eqs. (\ref{eq:Fterms}). Therefore, from downward and upward SUSY transformations inside the bracket of $(\frac{W}{M_P^2}\psi_\mu\sigma^{\mu\nu})\psi_\nu$, we obtain
$\sim \partial^\mu\lambda_{-1/2}\psi_\mu\sim (M_P/F)\partial^\mu\lambda_{1/2}\psi_\mu$,
\begin{equation}
\frac{F^2}{M_P^2}\partial_\mu\left(\frac{\lambda_{-1/2}}{\sqrt3}\right)
\sigma^{\mu\nu}\psi_\nu=\frac{F}{\sqrt3 M_P}(\partial_\mu\lambda_{G})
\sigma^{\mu\nu}\psi_\nu
\end{equation}
The SUSY breaking scale is $F=\sqrt{F_\Phi^2+F_{\Phi'}^2}=f^2[(-\log\xi+\log\epsilon)^2+\xi ^2]^{1/2}$. Therefore, the coefficient of  $(\partial^\mu \lambda_{G} ) \psi_\mu$  is
\begin{align}
 m_{3/2}^{\rm mixing}=\frac{f^2}{\sqrt3 M_P}[(-\log\xi+\log\epsilon)^2+\xi^2 ]^{1/2}\label{eq:gravmSUSYtrans}
\end{align}
where we labeled the mass as `mixing' since we calculated it from the mixing of spin-$\frac32$ and spin-$\frac12$ components. We can see that Eq. (\ref{eq:gravmSUSYtrans}) is basically the same as $ m_{3/2}$ of Eq. (\ref{eq:tunedgrm}).

Note that the vacuum energy is given by
\begin{align}
V_{\rm vac}&=k|V_{\rm vac}|=F^2 -3\frac{|W|^2}{M_P^2}\\
 &=\int  d^2\vartheta W -3\frac{|W|^2}{M_P^2}
 \end{align}
where $k=1,0$, and $-1$ for the dS, flat and AdS spaces, respectively. Namely,
\begin{align}
 \int  d^2\vartheta W&=\sum_{\phi_\kappa=\Phi,\Phi'} \frac{\partial W}{\partial \phi_\kappa}{F_{\phi_\kappa}}=3\frac{|W|^2}{M_P^2}+k|V_{\rm vac}|,
\end{align}
and the gravitino mass in the curved space has the same form in terms of $F$, i.e. $3\frac{|W|^2}{M_P^2}$ does not cancel $F^2$ exactly in the curved space.





\end{document}